\documentclass[pra,twocolumn,showpacs,preprintnumbers,amsmath,amssymb,superscriptaddress]{revtex4}
\usepackage{graphicx,amsmath,amssymb,amsfonts,latexsym,color,bm,epsfig,multirow,dsfont,verbatim}

\def\bbm[#1]{\mbox{\boldmath $#1$}}
\newcommand{\ket}[1]{\displaystyle{|#1\rangle}}
\newcommand{\bra}[1]{\displaystyle{\langle #1|}}
\newcommand{\cket}[1]{\displaystyle{|#1)}}
\newcommand{\cbra}[1]{\displaystyle{(#1|}}
\newcommand{\sumpk}{\sum_p\int_0^{+\infty}dk_z\int d^2\mathbf{k}}
\newcommand{\sumnb}{\sum_{n=1}^{+\infty}\sum_{b=1}^{+\infty}}
\newcommand{\apk}{a_p(\mathbf{k},k_z)}
\newcommand{\acpk}{a^\dag_p(\mathbf{k},k_z)}

\newcommand{\rv}{\mathbf{r}}

\newcommand{\Alkzt}{A_p^{(\theta)}(\mathbf{k},k_z,\mathbf{r})}

\newcommand{\Aclkztp}{A_p^{(\theta)\dag}(\mathbf{k},k_z,\mathbf{r'})}
\newcommand{\sumk}{\int_0^{+\infty}dk_z\int d^2\mathbf{k}}
\DeclareMathOperator{\Rea}{Re}
\DeclareMathOperator{\Ima}{Im}

\begin{document}

\title{Lifetimes of atoms trapped in an optical lattice in proximity of a surface}
\author{Sophie Pelisson}\email{sophie.pelisson@obspm.fr}\affiliation{LNE-SYRTE, Observatoire de Paris, CNRS UMR8630,
UPMC\\61 avenue de l'Observatoire, 75014 Paris, France}
\author{Riccardo Messina}\affiliation{LNE-SYRTE, Observatoire de Paris, CNRS UMR8630,
UPMC\\61 avenue de l'Observatoire, 75014 Paris, France}
\affiliation{Laboratoire Charles Fabry, Institut d'Optique, CNRS, Universit\'e Paris-Sud,
Campus Polytechnique, RD128, F-91127 Palaiseau Cedex, France}
\author{Marie-Christine Angonin}\affiliation{LNE-SYRTE, Observatoire de Paris, CNRS UMR8630,
UPMC\\61 avenue de l'Observatoire, 75014 Paris, France}
\author{Peter Wolf}\affiliation{LNE-SYRTE, Observatoire de Paris, CNRS UMR8630,
UPMC\\61 avenue de l'Observatoire, 75014 Paris, France}

\date{\today}

\begin{abstract}
We study the lifetime of an atom trapped in an optical vertical lattice in proximity of a massive surface using a complex scaling approach. We analyze how the presence of the surface modifies the known lifetimes of Wannier-Stark states associated to Landau-Zener tunnelling. We also investigate how the existence of a hypothetical short-distance deviation from Newton's gravitational law could affect these lifetimes. Our study is relevant in order to discuss the feasibility of any atomic- interferometry experiment performed near a surface. Finally, the difficulties encountered in applying the complex-scaling approach to the atom-surface Casimir-Polder interaction are addressed.
\end{abstract}


\maketitle

\section{Introduction}

The problem of atoms trapped in an accelerated periodic lattice is an old one in solid state physics \cite{Ashcroft,WannierPhysRev60,MoiseyevPhysRep98,GluckEurJPhysD98,GluckPhysRep02}. However, some phenomena like pseudo-eigenstates (Wannier-Stark ladders) and the lifetime limitations of those states (Landau-Zener tunneling) are still a current subject of research both on the experimental side and theoretically \cite {WimbergerJPhysB06,WimbergerPRA05,SiasPRL07,ZenesiniPRL09,TayebiradPRA10,LorchPRA12,WitthautPRA07,RapediusPRA10} (and see \cite{GluckPhysRep02} and reference therein for a more extensive review). In this work, we study the Landau-Zener effect \cite{LandauPhysZSowjetunion32,ZenerProcRoySocLond32,ZenerRSocLond34} for atoms trapped in a vertical standing wave in the presence of short scale interactions.

This is done in the context of the theoretical modeling of the experiment FORCA-G (FORce de CAsimir et Gravitation \`{a} courte distance) aiming at measuring the Casimir-Polder force acting on a neutral atom in the presence of a massive dielectric surface by atomic interferometry. At the same time, the experiment is configured to search for non-Newtonian deviations from the gravitational law at short distances \cite{LemondePRA05,WolfPRA07,MessinaPRA11, BeaufilsPRL11,TackmannPRA11,PelissonPRA12,PellePRA13}. These two goals are reached by the use of atoms trapped in a vertical optical standing wave in front of a massive surface realizing the trap. Such a lattice allows a precise knowledge of the distance between the atom and the surface, this distance being a multiple of the wavelength of the trap. The atomic states in such a trap were the main subject of a previous paper \cite{MessinaPRA11}. In that paper the shape of the wavefunctions and the modification of energy levels due to the presence of the surface were deduced by solving a standard eigenvalue problem. In this context, the states are supposed to be bound states of the system, meaning that their lifetime is infinite.

Nevertheless, solid state physics teaches us that the states in an accelerated periodic lattice are not bound states but resonance states with a finite lifetime. The subject of this paper is to quantify these lifetimes for the states calculated in \cite{MessinaPRA11} in order to check the feasibility of the experiment, knowing that the measurement of the interaction by atom interferometry takes at most a few seconds. We want to understand, more in detail, how the presence of the surface and of a hypothetical non-Newtonian gravitational potential modifies the well-known finite lifetimes of standard Wannier-Stark states.

This paper is organized as follows. In Sec. \ref{Sec:2}, we describe the system under scrutiny and the method we use to calculate the lifetime of the atomic states. Section \ref{Sec:3} is devoted to the effect of the surface. This section is separated in two parts. As a first step, we study the effect of a surface treated as a boundary condition for an atom below it. In the following, we investigate the phenomena arising when tunneling of the atom through the surface is allowed. This is done for an atom above the surface in order to clearly discriminate the effect of the finiteness of the potential barrier from other resonance effects. Then, Sec. \ref{Sec:4} shows the effect of a deviation from Newton's gravitational law at short distance on the lifetimes of the trapped atom whereas Sec. \ref{Sec:5} discusses the problem arising when the Casimir-Polder interaction is taken into account. In this part, the difficulties of the treatment of the Casimir-Polder effect by the complex-scaling method are highlighted.

\section{Lifetimes of atoms in Wannier-Stark states}\label{Sec:2}
Let us consider the Hamiltonian of an atom of mass $m_a$ trapped in a periodic lattice $V(z)=\frac{U}{2}\left(1-\cos(2k_lz)\right)$ in the presence of a linear gravitational potential
\begin{equation}
H_{\footnotesize{\text{{WS}}}}=\frac{p^2}{2m_a}+\frac{U}{2}\left(1-\cos(2k_lz)\right)-m_agz.
\label{hamws}
\end{equation}
where $U$ is the depth of the trapping potential, $k_l$ represents the wavevector of the trap and $g$ is the earth gravitationnal constant.
This operator is the sum of the well-known Bloch Hamiltonian $H_{B}=\frac{p^2}{2m_a}+\frac{U}{2}\left(1-\cos(2k_lz)\right)$ \cite{Ashcroft} and a linear potential $-m_agz$ (the $z$ axis is oriented downwards). As a consequence, the eigenstates of Eq. \eqref{hamws} are derived from the Bloch states, generally taking into account one single Bloch band. The spectrum obtained using this procedure can be written as
\begin{equation}
E_{\alpha,n}=\bar{\varepsilon}_\alpha-nm_ag\frac{\lambda_l}{2},
\label{enws}
\end{equation}
where $\frac{\lambda_l}{2}$ is the periodicity of the lattice, $\lambda_l$ being its wavelength, $\bar{\varepsilon}_\alpha$ is the mean value of the energy of the Bloch band $\alpha$ under consideration and $n$ is an integer labelling the well where the state is centered. The states obtained are known as Wannier-Stark states and were first discussed for electrons in a crystal submitted to a constant electric field \cite{WannierPhysRev60}. The existence of such eigenstates results from the single-band approximation. Nevertheless, a calculation ignoring this approximation confirms the existence of these states, but associates a finite lifetime to each of them. This lifetime physically corresponds to the so-called Landau-Zener tunneling, namely the possibility for a particle in a Wannier-Stark lattice site to transist from one Bloch band to another \cite{LandauPhysZSowjetunion32,ZenerProcRoySocLond32,ZenerRSocLond34}. In other words, the discrete spectrum in Eq. \eqref{enws} is immersed in the continuum of eigenvalues of the Hamiltonian \eqref{hamws} and can thus be seen as a resonance spectrum \cite{AglerCommunMathPhys85}. The spectrum of Eq. \eqref{enws} should thus be rewritten as \cite{MoiseyevPhysRep98,GluckEurJPhysD98,GluckPhysRep02}
\begin{equation}
E_{\alpha,n}=\bar{\varepsilon}_\alpha-nm_ag\frac{\lambda_l}{2}-i\frac{\Gamma_\alpha}{2}.
\label{enres}
\end{equation}
The eigenstates associated to the spectrum \eqref{enres} are metastable states with a finite lifetime $\tau=\frac{\hbar}{\Gamma_\alpha}$. In order to completely characterize the states of an atom in our optical lattice, we need to work out this complex spectrum to evaluate their lifetime in the trap.

Several methods have been developed so far in order to study this complex spectrum \cite{AvronPRL76,BentoselaJPhysC82,GrecchiJPhysA93,GrecchiCommunMathPhys97,BuslaevJMathPhys98,GlutschPRB99}, each using approximations such as a periodic potential with a finite number of gaps or a finite periodic lattice. Most of these methods are based on the analogy of the Wannier-Stark sates with scattering states. In this theoretical framework, the complex spectrum \eqref{enres} can be calculated as the set of poles of the scattering matrix of the system. The method we will use in this work is based on the rotation of the Hamiltonian \eqref{hamws} in the complex plane and the derivation of the complex spectrum as eigenvalues of the obtained non-hermitian Hamiltonian. This approach is known as \emph{complex-scaling} method \cite{MoiseyevPhysRep98,ReinhardtAnnRevPhysChem82,BuchleitnerJPhysB94} and the rotation of the Hamiltonian is performed via the operator $C(\theta)$ such that \begin{equation}
C(\theta)\varphi(z)=\varphi(ze^{i\theta}) \qquad \forall\varphi\in C_0^\infty.
\label{scaling}
\end{equation}
The one-dimensional complex-scaled Hamiltonian is then
\begin{equation}
H_{\footnotesize{\text{WS}}}^{(\theta)}=\frac{e^{-2i\theta}p^2}{2m_a}+\frac{U}{2}\bigl(1-\cos(2k_lze^{i\theta})\bigr)-m_agze^{i\theta}.
\label{hamcs}\end{equation}
According to the Balslev-Combes theorem \cite{BalslevCommunMathPhys71}, the complex spectrum \eqref{enres} can then be found by solving the eigenvalue problem
\begin{equation}
H_{\footnotesize{\text{WS}}}^{(\theta)}\ket{\phi_{\alpha,n}}=E_{\alpha,n}\ket{\phi_{\alpha,n}},
\label{Schro}\end{equation}
where the real part of a given eigenvalue represents its energy level, while its imaginary part is the width of the resonance.

\begin{figure}\center
\includegraphics[height=4cm]{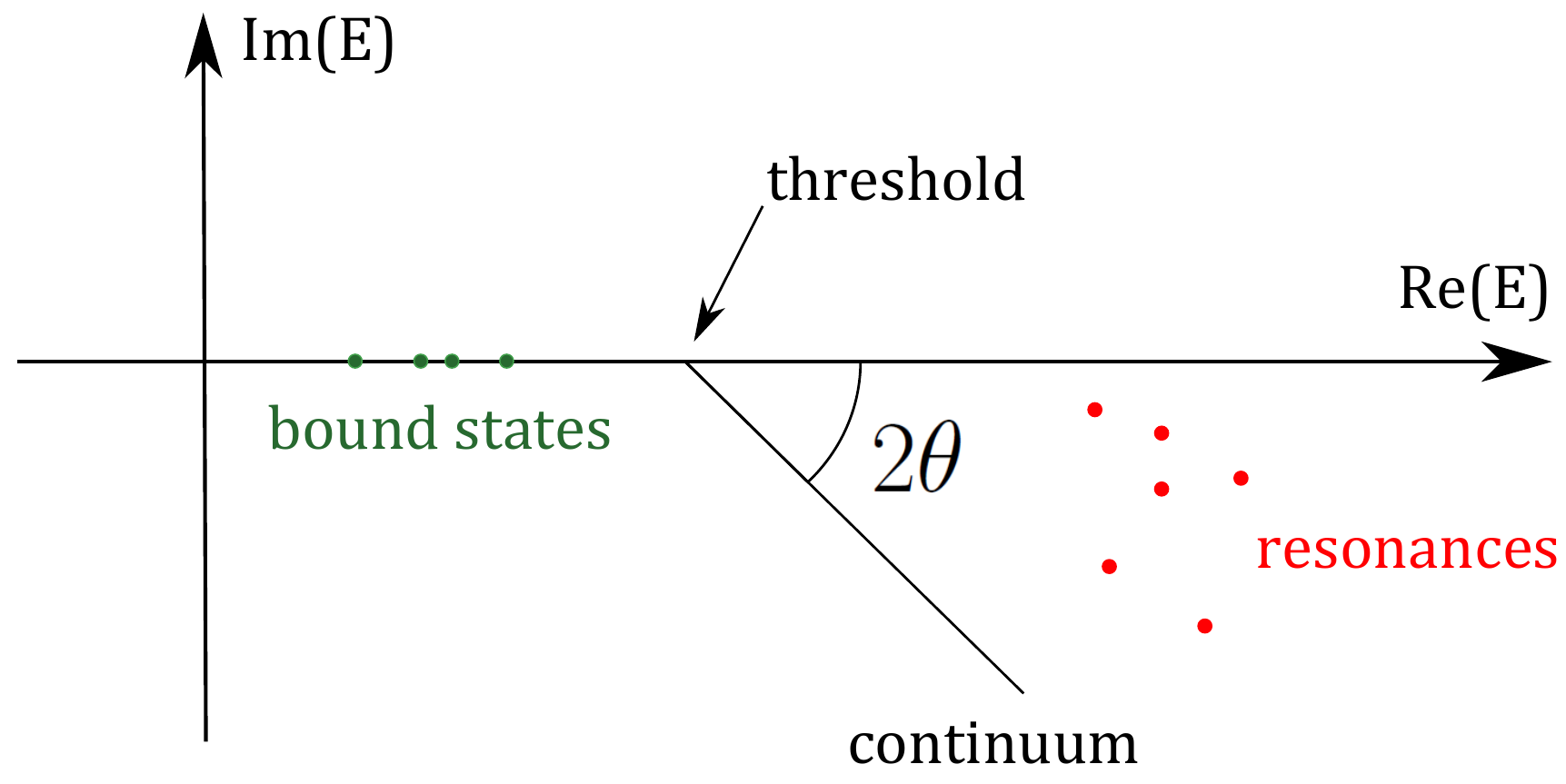}
\caption{(Color online) Schematic representation of the complex eigenvalues of the complex-coordinate scaled Hamiltonian $H^{(\theta)}$ according to the Balslev-Combes theorem \cite{BalslevCommunMathPhys71}.}
\label{EV}
\end{figure}

\begin{figure}\center
\includegraphics[height=5.3cm]{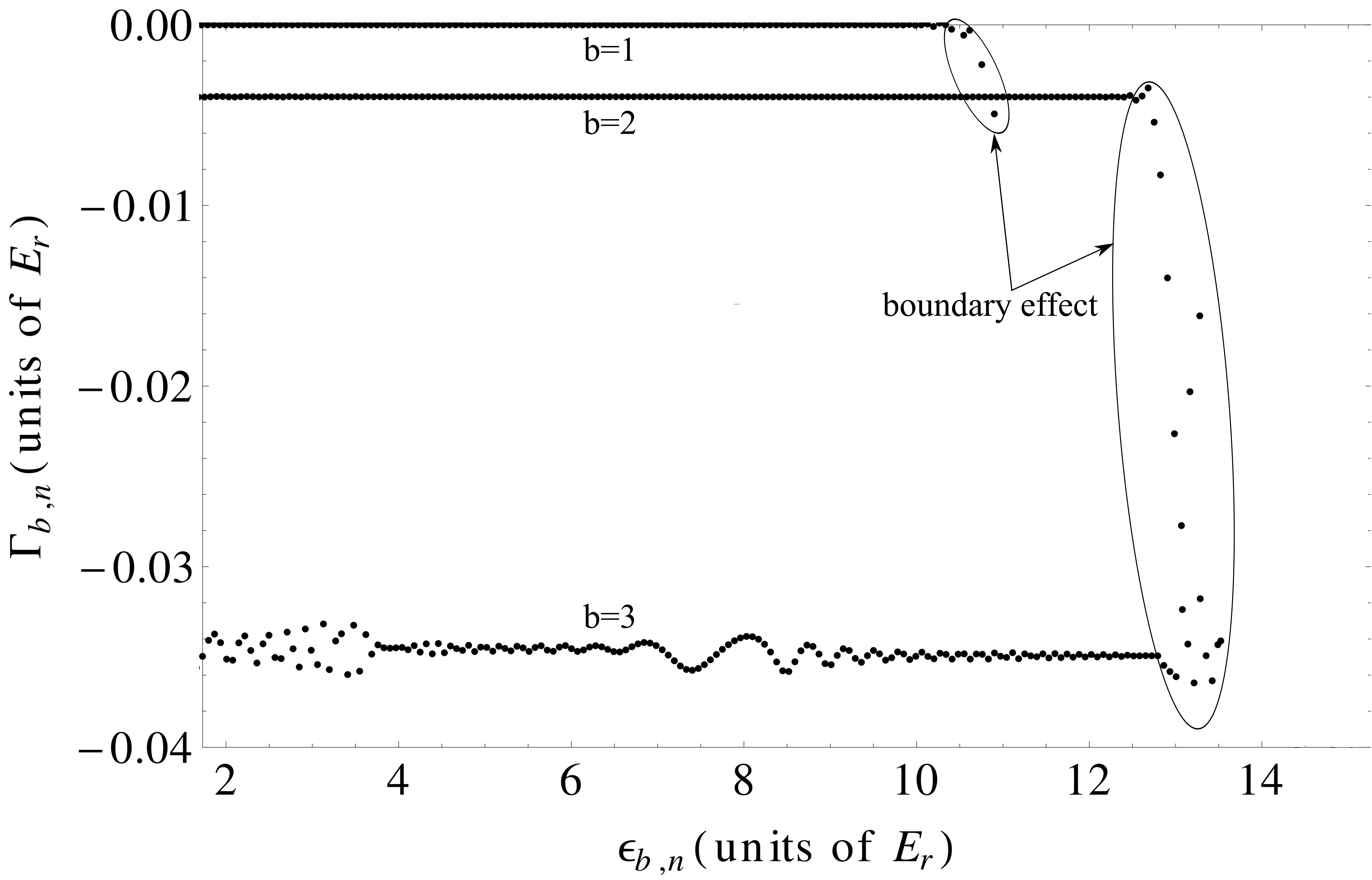}
\caption{Complex eigenvalues for the 3 first Bloch bands of the
Hamiltonian $H_{\footnotesize{\text{WS}}}^{(\theta)}$. The
energies are expressed in units of $E_r=\frac{\hbar^2k_l^2}{2m_a}$
which is the natural unit for energy levels in a trap with
wavevector $k_l$ (in the case of FORCA-G,
$E_r\simeq5.37\cdot10^{-30}$\,J and $\Gamma=1E_r$ corresponds to
$\tau=\hbar/\Gamma=1.96\cdot10^{-5}$\,s). The spectrum here is calculated for $u=3E_r$.} \label{EVWS}
\end{figure}

We stress here that resonance wavefunctions are characterized by a divergent behavior when $z\to\infty$ as
\begin{equation}\phi_{res}(z\to\infty)\simeq e^{-ikz}+S(k)e^{ikz},\label{div}\end{equation}
$k$ being the wavevector of the atom and $S(k)$ the scattering matrix of the system. As demonstrated in \cite{MoiseyevPhysRep98}, the divergent part of the resonant wavefunctions in Eq. \eqref{div} is regularized by the complex scaling. This is not the case for a non-resonant wavefunction of the continuum. As a consequence, it is important to distinguish the resonant from the non-resonant part of the complex spectrum worked out solving Eq. \eqref{Schro}. A schematic representation of the eigenvalues of Hamiltonian \eqref{hamcs} is shown in Fig. \ref{EV}.
We see in this figure that the continuum eigenvalues lie on a line forming an angle $2\theta$ with the real axis, $\theta$ being the angle chosen for the transformation \eqref{scaling}. At the same time the real bound states are located on the real axis (corresponding to the fact that they have an infinite lifetime) while the resonance states are the complex eigenvalues which are not part of the rotated continuum. In the case of the Wannier-Stark Hamiltonian, the presence of energy bands due to the periodicity of the trap leads to a peculiar structure of the complex eigenvalues. The eigenvalues of the Hamiltonian \eqref{hamcs} are shown in Fig. \ref{EVWS}.
These eigenvalues are obtained using the pseudo-periodicity of the trap. Indeed, the fact that we calculate states in an infinite pseudo-periodic lattice allows us to restrict the search of the spectrum in Eq. \eqref{enres} to the analysis of the spectral properties of the Floquet-Bloch operator \cite{GluckPhysRep02}
\begin{equation}
U=e^{-\frac{i}{\hbar}H_{\footnotesize{\text{WS}}}T_B}\hspace{1cm} T_B=\frac{2\hbar}{m_ag\lambda_l},
\end{equation}
over a Bloch period \cite{GluckEurJPhysD98,GluckPhysRep02}. The eigenvalues are then worked out using a Discrete Variable Representation (DVR) method and a QR algorithm for complex symmetric matrices \cite{Press95}.

In order to check the validity of our method, we compare the real parts of the calculated eigenvalues with the spectrum obtained using two different methods. The first is the tight-binding model \cite{GluckPhysRep02,LemondePRA05}, a standard semi-analytical resolution of the Schr\"odinger equation for $H_{\footnotesize{\text{WS}}}$ which does not give access to the finite lifetime of the states. The second one is a finite-difference numerical approach to $H_{\footnotesize{\text{WS}}}$ without complex scaling. Table \ref{valpws} shows the good agreement between the real parts of eigenvalues \eqref{enres} and the second approach. The difference between our results and the tight-binding model can be traced back to the fact that this approach restricts the Hilbert space to the first Bloch band.
\begin{center}\begin{table}[h]\begin{tabular}{|c|c|c|c|}
\hline
$n$ & $\Rea(E_{1,n})$ & $\mathcal{E}_{1,n}$ & $\mathcal{E}_n$\\
\hline\hline
-5  & 1.78711 & 1.78711 & 1.78718 \\
-4  & 1.71703 & 1.71703 & 1.71711 \\
-3  & 1.64696 & 1.64696 & 1.64704 \\
-2  & 1.57688 & 1.57688 & 1.57697 \\
-1  & 1.50681 & 1.50681 & 1.50690 \\
0   & 1.43674 & 1.43674 & 1.43683 \\
1   & 1.36667 & 1.36667 & 1.36677 \\
2   & 1.29660 & 1.29660 & 1.29670 \\
3   & 1.22652 & 1.22652 & 1.22663 \\
4   & 1.15645 & 1.15645 & 1.15656 \\
5   & 1.08638 & 1.08638 & 1.08649 \\
\hline\end{tabular}\caption{Table of the real parts $\Rea(E_{1,n})$ of the complex energies \eqref{enres} compared with the Wannier-Stark spectrum $\mathcal{E}_{1,n}$ obtained numerically without complex scaling and with the energies $\mathcal{E}_n$ resulting from a semi-analytical calculation in the first Bloch band. The energies are given in units of $E_r=\frac{\hbar^2k_l^2}{2m_a}$ which is the natural unit for energy levels in a trap with a wavevector $k_l$. The numerical example corresponds to $U=3E_r$.}\label{valpws}\end{table}\end{center}

Concerning the imaginary parts, representing the lifetime of the band under scrutiny, Niu et al. have proposed a general formula to evaluate the inverse lifetime of the standard Wannier-Stark states \cite{NiuPRL96}. This formula gives an estimate of the tunneling rate $\gamma$ as a function of a critical acceleration $\alpha_c$ and the dimensionless accelaration $\alpha$ due to the linear potential.
\begin{equation}
\gamma=\alpha e^{-\alpha_c/\alpha}.
\end{equation}
In our case, $\alpha=\frac{m_a^2g}{\hbar^2k_l^3}$ and the critical acceleration can be expressed as
\begin{equation}
\alpha_c=\frac{\pi\Delta^2}{K},
\end{equation}
where $\Delta$ is the half width of the energy gap between the first and the second Bloch band and $K=\frac{n}{2}$ is the wave number of Bragg scattering corresponding to the $n$-th gap (here we will take $n=1$). With this expression, we obtain, for the second Bloch band, a tunneling rate of the order of $0.003\,E_r$ which is of the same order of magnitude of the value of the complex part of the spectrum \eqref{enres} as it can be seen on Fig. \ref{EVWS}. An additional verification with other values of the well depth assures us that the complex-scaling method provides satisfying results for this physical system. As a consequence, in the next section, we will use this method to analyze the lifetimes of the Wannier-Stark states in the presence of a surface, whose main features have been already discussed in \cite{MessinaPRA11}.

\section{Lifetimes in the presence of the surface}\label{Sec:3}

\subsection{Presence of the surface}

In our problem, we consider the presence of a surface at $z=0$. Our atom is thus located below the surface, in the region $z>0$ (we recall that the $z$ axis is oriented downwards). The presence of this boundary condition breaks the quasi-periodicity of the system. The potential modifying the optical trap is no longer linear, since it must be considered as the gravitational linear potential for $z>0$ and an infinite potential barrier for $z\leq0$, describing the impossibility of the particle to penetrate the mirror. We have shown in a previous paper \cite{MessinaPRA11} that this surface induces a modification of the energies and states of the Hamiltonian $H_{\footnotesize{\text{WS}}}$. We now want to verify that this modification does not reduce drastically the lifetimes of our atoms in the trap.

As anticipated before, the presence of a boundary condition in
$z=0$ breaks the translational symmetry of the problem. Thus, we
can no longer use the quasi-periodic approximation and we have to
solve directly the eigenvalues problem \eqref{Schro}. This is done
using a finite-element method and once again a QR algorithm for
symmetric complex matrices using the subroutine \verb!zgeev! of
the numerical package \verb!LAPACK! \cite{Anderson99}. However,
using this method, numerical problems arise when the well depth is
larger than $U=2.2E_r$, since in this case the imaginary parts of
the complex eigenvalues are too small. As a matter of fact, the
lifetime of a given Wannier-Stark well is a strongly increasing function of the well depth. This dependence can be well
understood using the image of tunneling. Indeed, the finite
lifetime of the metastable states can be seen as a consequence of
the possibility for an atom in a lattice site $n$ to tunnel
through the potential toward an upper Bloch band in another site.
This is favorized by the linear potential which induces resonant
tunneling between the first band in a given site to e.g. the
second band in a farther site. As a consequence the atom reaches a
new state with a new energy destroying the former one. Moreover,
the atom can reach a state with an energy higher than the trap
depth, being thus no longer trapped. So, the tunneling probability
depends on the lattice depth: the deeper the trap is, the weaker
this probability becomes.

\begin{figure}\center
\includegraphics[height=5.7cm]{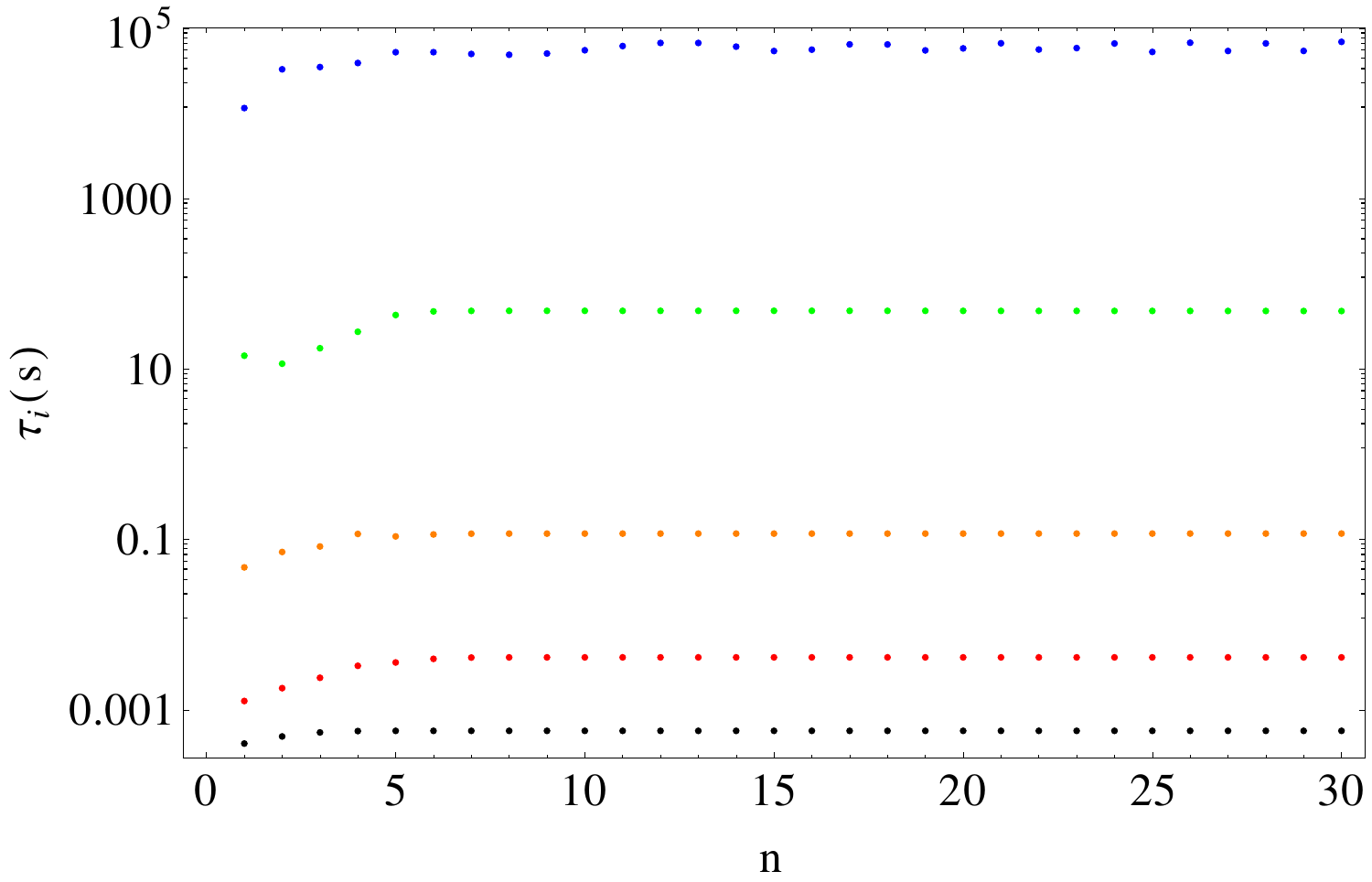}
\caption{(Color online) Lifetimes of the first Bloch band in function of the quantum number $n$ labelling the well of the trap for five different depths. From top to bottom we have: $U=2E_r$ (blue dots), $U=1.5E_r$ (green dots), $U=1E_r$ (orange dots), $U=0.5E_r$ (red dots), $U=0.1E_r$ (black dots).}
\label{bandtau}
\end{figure}

This feature is illustrated in Fig. \ref{bandtau}, where lifetimes are calculated for different well depths.
Here we have taken into account only the lifetime of the first Bloch band. Indeed, as it can be seen in Fig. \ref{EVWS}, the lifetime of the first band is the longest one, and the larger the band under scrutiny is, the shorter is this lifetime. In the particular case of our experimental parameters, the second Bloch band has a mean energy of $\bar{\epsilon}_2\simeq5.45\,E_r$ and is then not trapped by a well depth $U=3E_r$. As a consequence, only the lifetime of the first band is of interest for us.

\begin{figure}\center
\includegraphics[height=5.3cm]{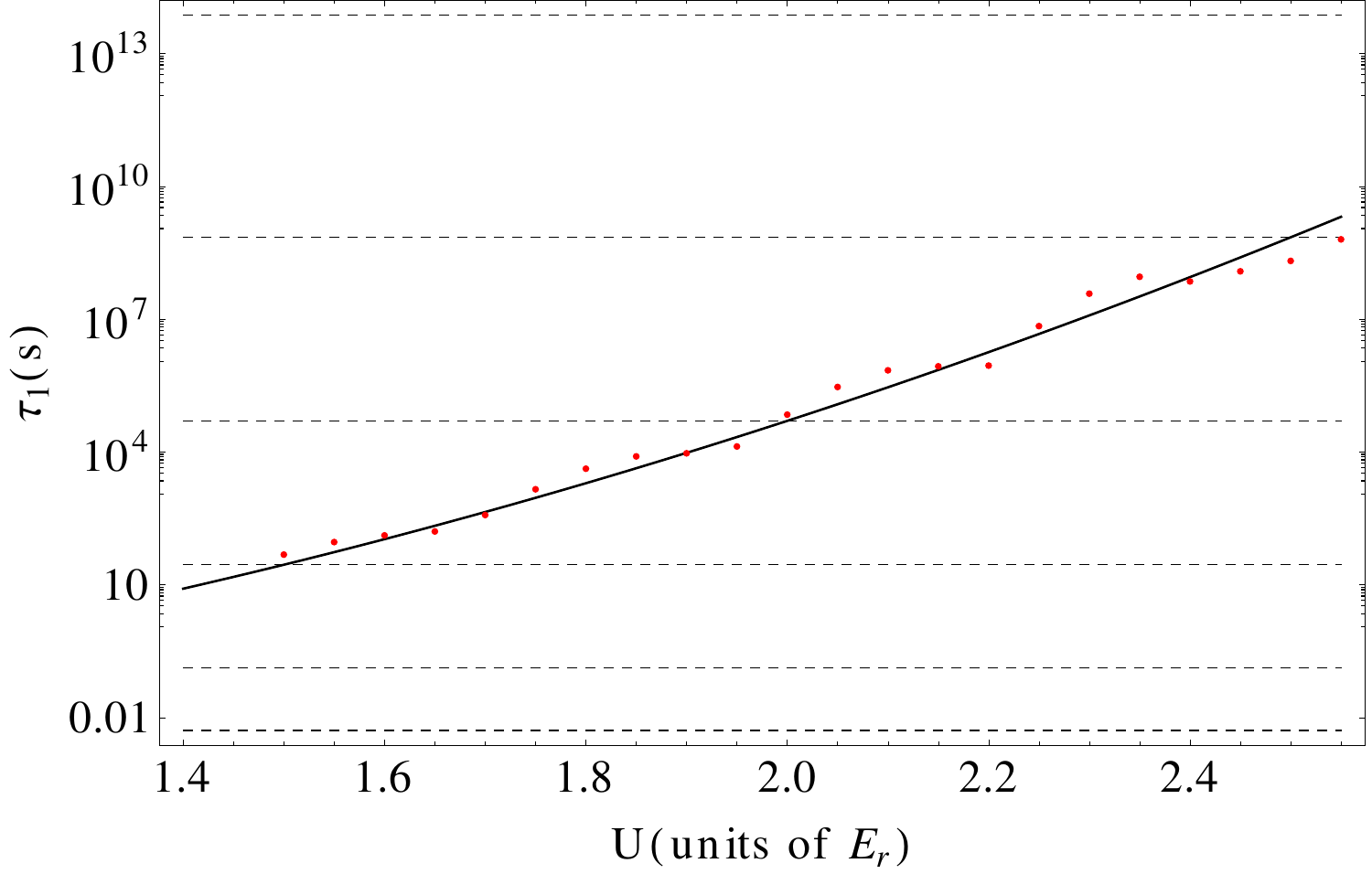}
\caption{(Color online) Lifetime of the first Bloch band calculated with the complex scaling method (Red dots) in function of the trap depth. The calculated lifetimes are compared with the Landau-Zener formula (Black line) for standard Wannier-Stark sates.}
\label{LZtau}
\end{figure}

The analysis of Fig. \ref{bandtau} shows, moreover, that for any
well depth, the first-band lifetimes are function of the
considered well. In particular, the lifetimes do not vary
remarkably starting approximately 5 wells away from the surface,
whilst being functions of the considered well in proximity of the
boundary $z=0$. In order to verify the coherence of the
calculation the constant lifetimes far from the surface (function
of the well depth) can be compared to the ones obtained through
the Landau-Zener formula, derived for the standard Wannier-Stark
states (i.e. in an infinite lattice). This comparison is shown in
Fig. \ref{LZtau}.
In this figure, we can see that the Landau-Zener formula still gives a good description far from the surface. We observe typical oscillations of lifetimes around the Landau-Zener results, already discussed theoretically in \cite{GluckPhysRep02} and observed experimentally in \cite{SiasPRL07}. These oscillations result from resonant tunneling occurring for specific values of $U$ between the most stable band and excited ones (for a detailed discussion see \cite{GluckPhysRep02}). This suggests that the lifetimes of our metastable states in the presence of a surface are of the order of $10^{14}\,$s for a well depth of $U=3E_r$. However, in this analysis we have treated the mirror as an infinite barrier at $z=0$, which is not very realistic. It is of interest to investigate how the results are modified if the mirror is modeled by means of a finite potential barrier. This will be done assuming that the atom is located above the mirror.

\subsection{Atom above the mirror}

\begin{figure}\center
\includegraphics[height=4.5cm]{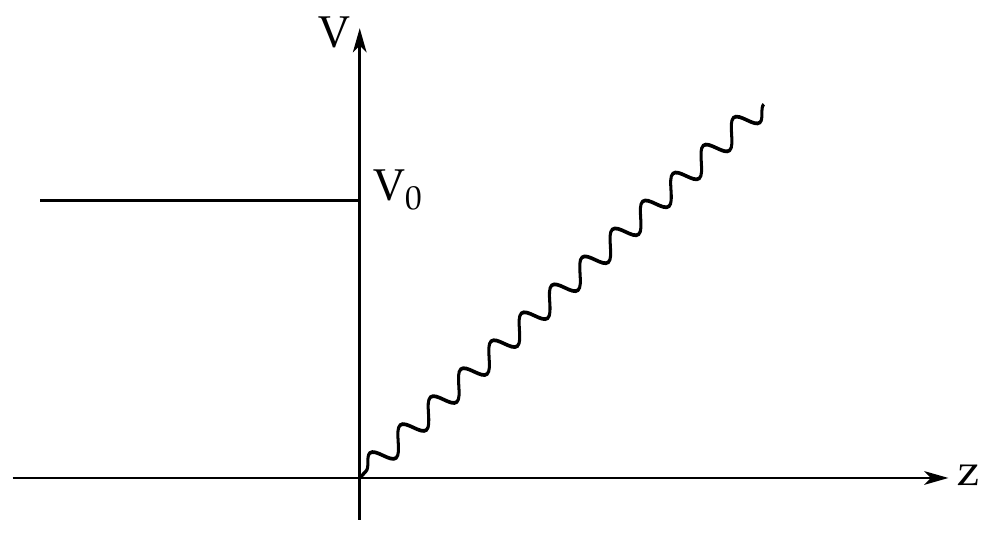}
\caption{Shape of the potential for atoms above a finite surface.}
\label{potabove}
\end{figure}

A way to increase the lifetime of the states is to place the atom above the surface rather than below. In this case, if we consider as before the mirror as an infinite barrier, the potential is a well of infinite height. As a consequence, it supports only a discrete set of real eigenstates having an infinite lifetime. In order to model the surface more precisely, we allow the tunneling of the atom into this surface. Therefore we replace the infinite barrier by a finite one as shown in Fig. \ref{potabove}.
As a consequence, the complex-scaled Hamiltonian of the system can be expressed as
\begin{equation}\begin{cases}H_{\footnotesize{\text{above}}}^{(\theta)}=\frac{e^{-2i\theta}p^2}{2m_a}+V_0\hspace{2cm}\text{if }z\leq0\\
H_{\footnotesize{\text{above}}}^{(\theta)}=\frac{e^{-2i\theta}p^2}{2m_a}+\frac{U}{2}\bigl(1-\cos(2k_lze^{i\theta})\bigr)\\
\hspace{1.5cm}+m_agze^{i\theta}\qquad\hspace{1.6cm}\text{elsewhere},\end{cases}\label{hamabove}\end{equation}
where $V_0\in\mathds{R}$.
In order to study the behavior of the atoms in front of a surface with a finite size, we investigate the complex spectrum for a given well depth ($U=1E_r$ in the numerical example) as a function of the height of the barrier chosen to represent the surface. More specifically, we focus our attention on the transition from bound states to resonances for atoms in front of a potential barrier with a varying height.

The resulting complex eigenvalues for different heights of the barrier are shown in Fig. \ref{above}.
\begin{widetext}\begin{center}\begin{figure}[h!]
\includegraphics[height=4.8cm]{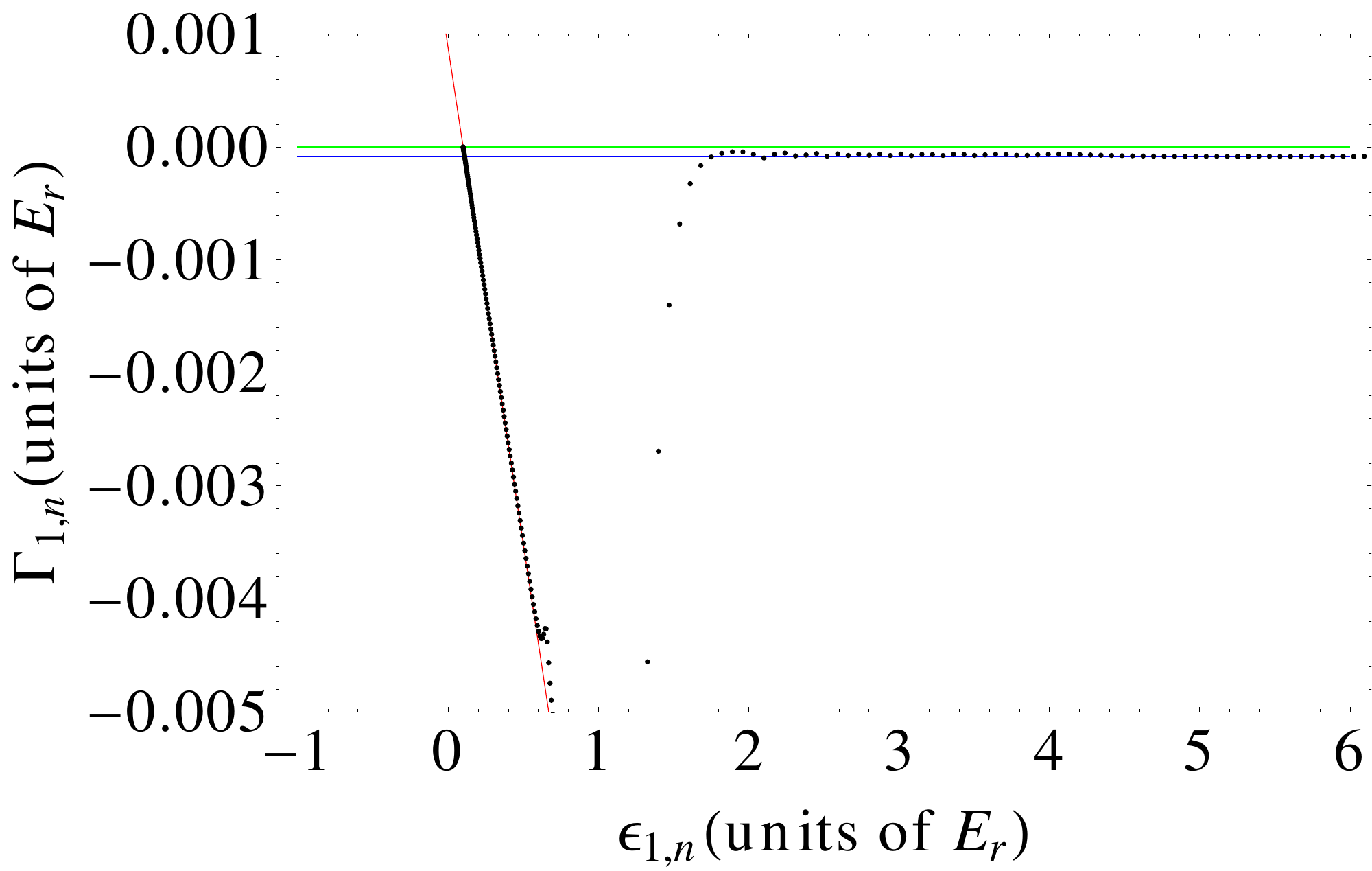}\hspace{1cm}\includegraphics[height=4.8cm]{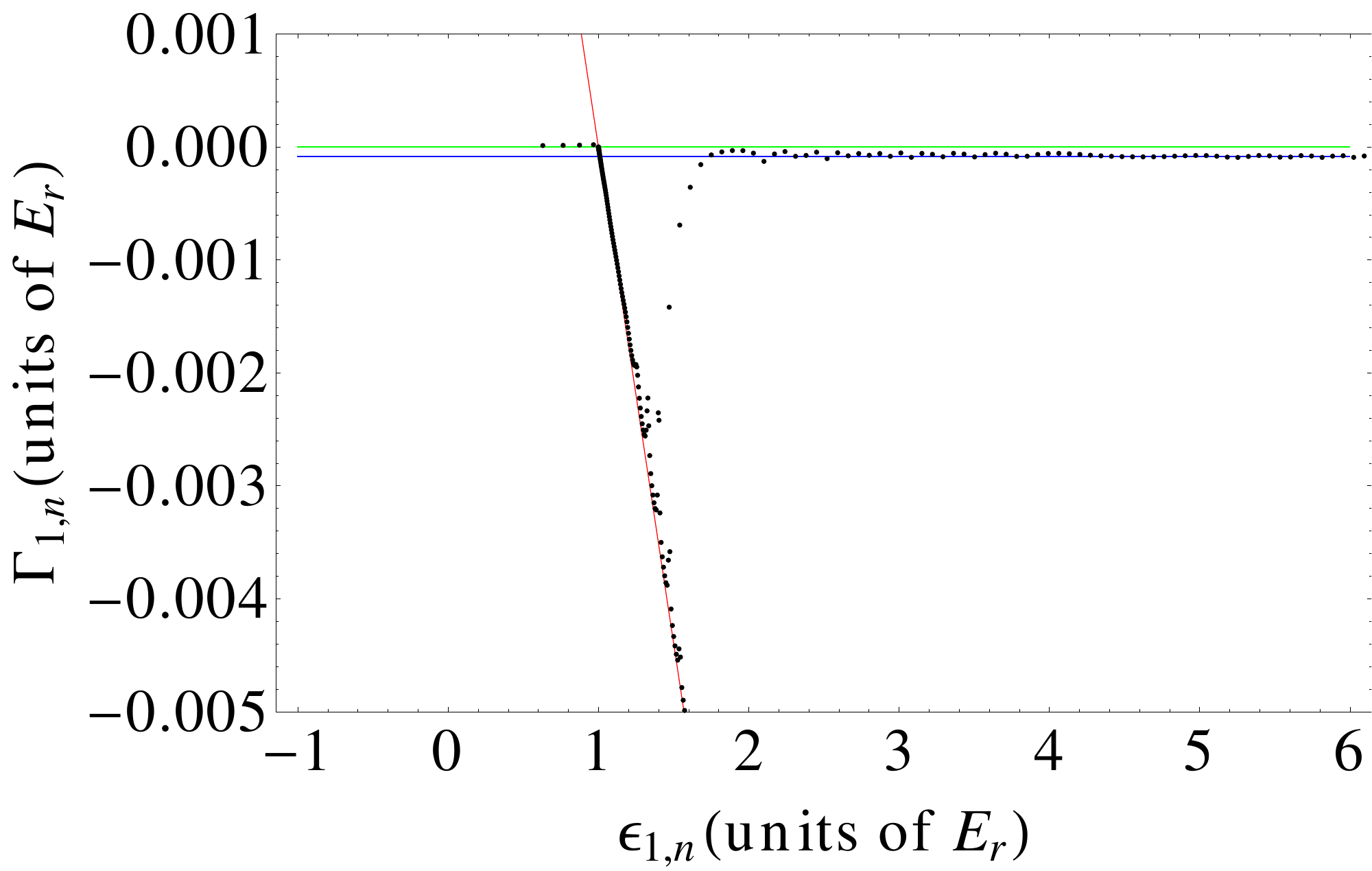}\\
\includegraphics[height=4.8cm]{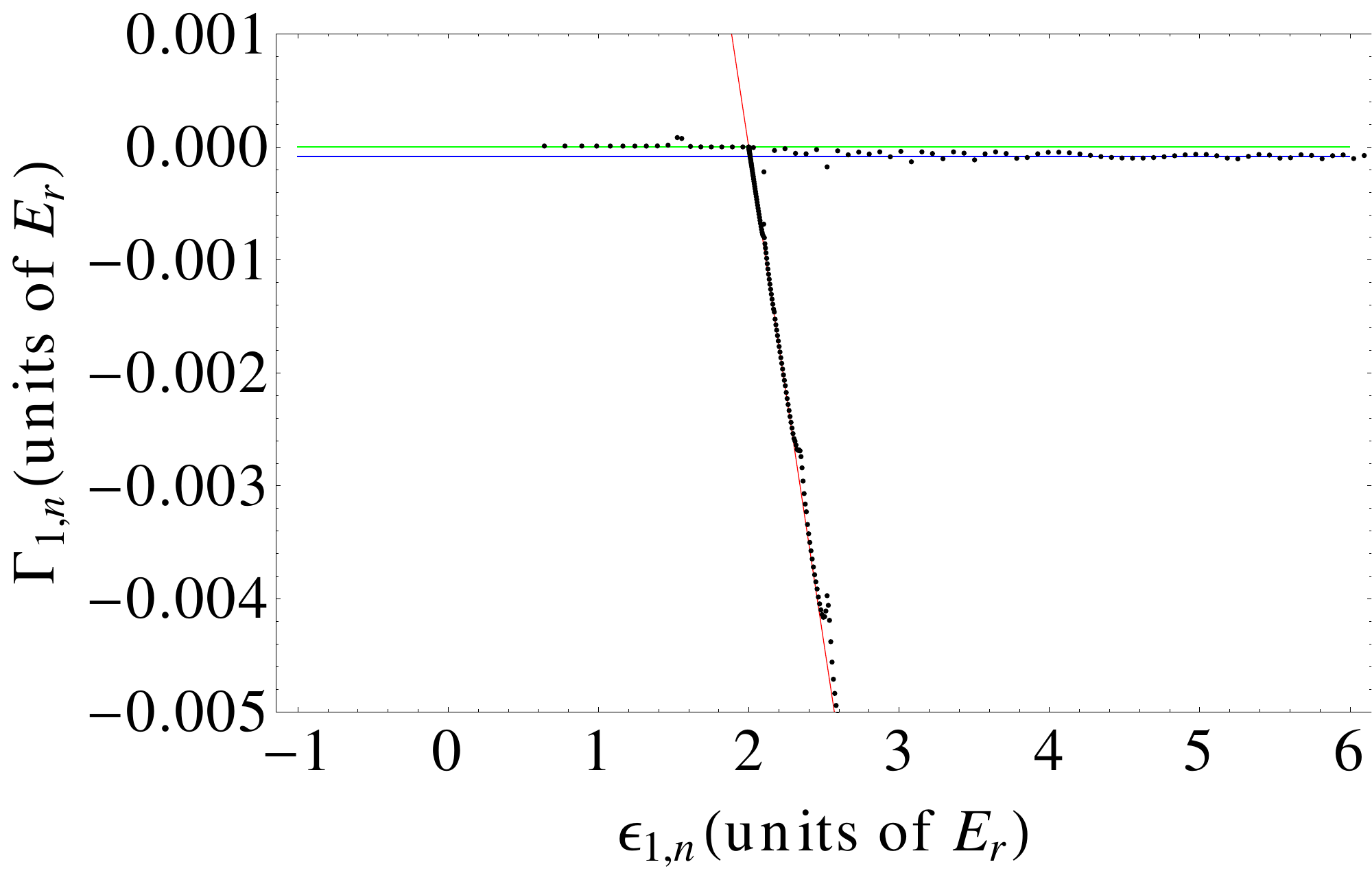}\hspace{1cm}\includegraphics[height=4.8cm]{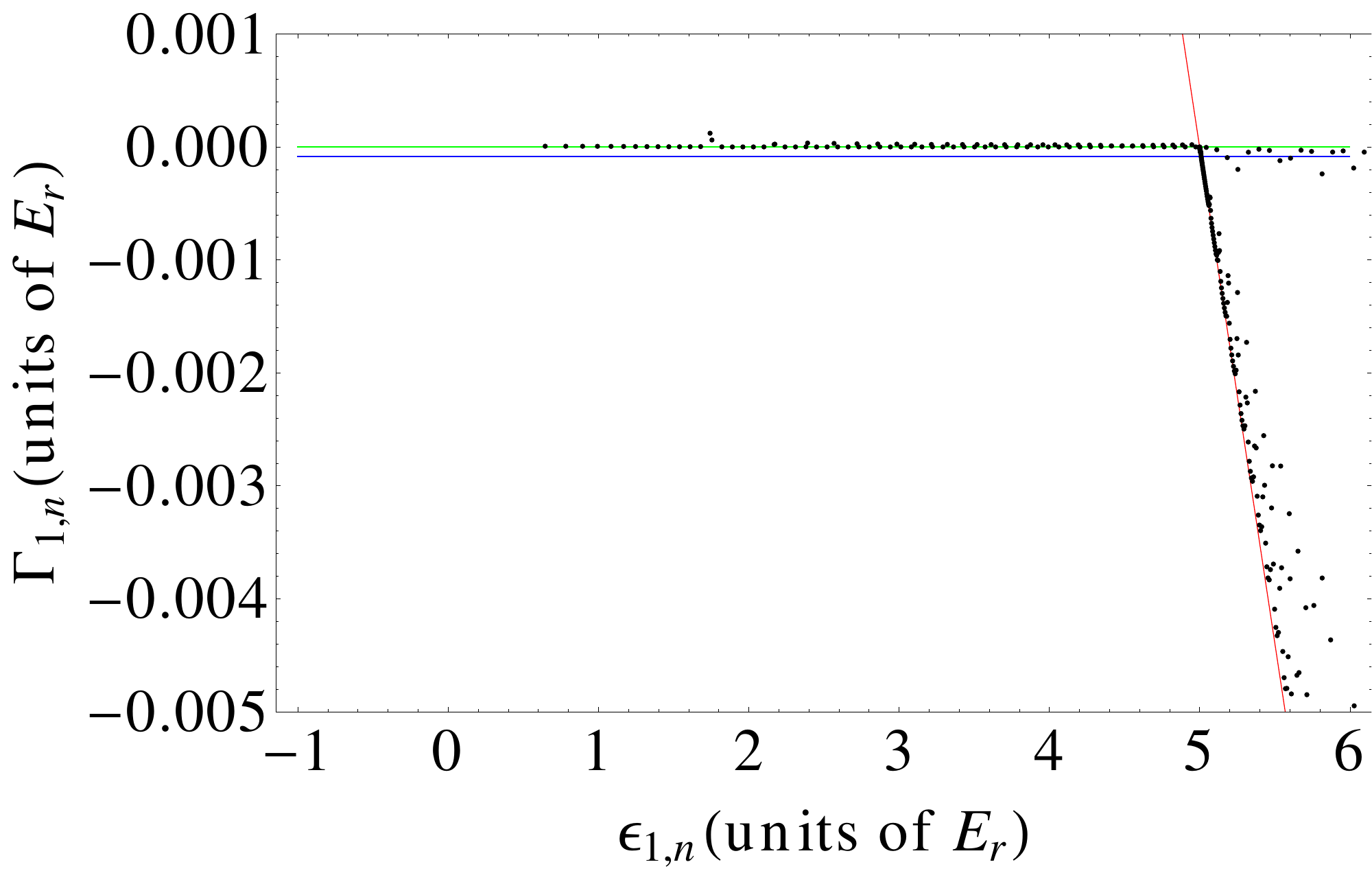}
\caption{(Color online) Complex spectrum of the Hamiltonian \eqref{hamabove} describing an atom in front of a potential barrier with different heights (from left to right and top to bottom $V_0=0.1E_r$, $V_0=1E_r$, $V_0=2E_r$ and $V_0=5E_r$) (black dots) in comparison with the standard Wannier-Stark lifetimes (blue lower horizontal line). The depth of the periodic trap is chosen to be $U=E_r$. The solid red line represent the line of the continuum with a slope of $2\theta$ and the horizontal green upper line highlights the eigenvalues with an imaginary part equals to 0.}
\label{above}
\end{figure}\end{center}\end{widetext}
In this figure, we see that the presence of the barrier induces the emergence of a continuum associated with the standard resonances spectrum. The position of this continuum depends on the height of the barrier: more precisely, the continuum region of the spectrum always starts at a value around $V_0$. When the barrier is very low (i.e. of the order of the depth of the periodic trap or below), we observe that the lifetime of the atom close of this barrier is shorter than the lifetime at longer distances. This can be understood thanks to the interpretation of Landau and Zener of the resonance phenomenon. Indeed, if we consider that the finite lifetime of the atom arises from the resonant tunnelling through the trap, it is clear that this tunnelling has a higher probability when the atom is close to the barrier because it is not trapped on the left side. In addition, we observe that when the barrier becomes higher than the well depth, some bound states appear corresponding to the closest states submitted to a potential well.

As expected, the effect of a finite barrier representing the surface is completely negligible for the wells far from the surface and starts to be visible for the closest wells when the barrier height is of the order of the well depth or below. This means that the penetrability of the surface should be comparable to that of the periodic trap in order to allow a significant modification of the behavior of the complex spectrum. This is reasonably not at all the case for a solid mirror and we can conclude that the finiteness of the surface should not play a major role in the lifetime of our states in the trap.

\section{Non-Newtonian gravitation}\label{Sec:4}

\begin{figure}\center
\includegraphics[height=7cm]{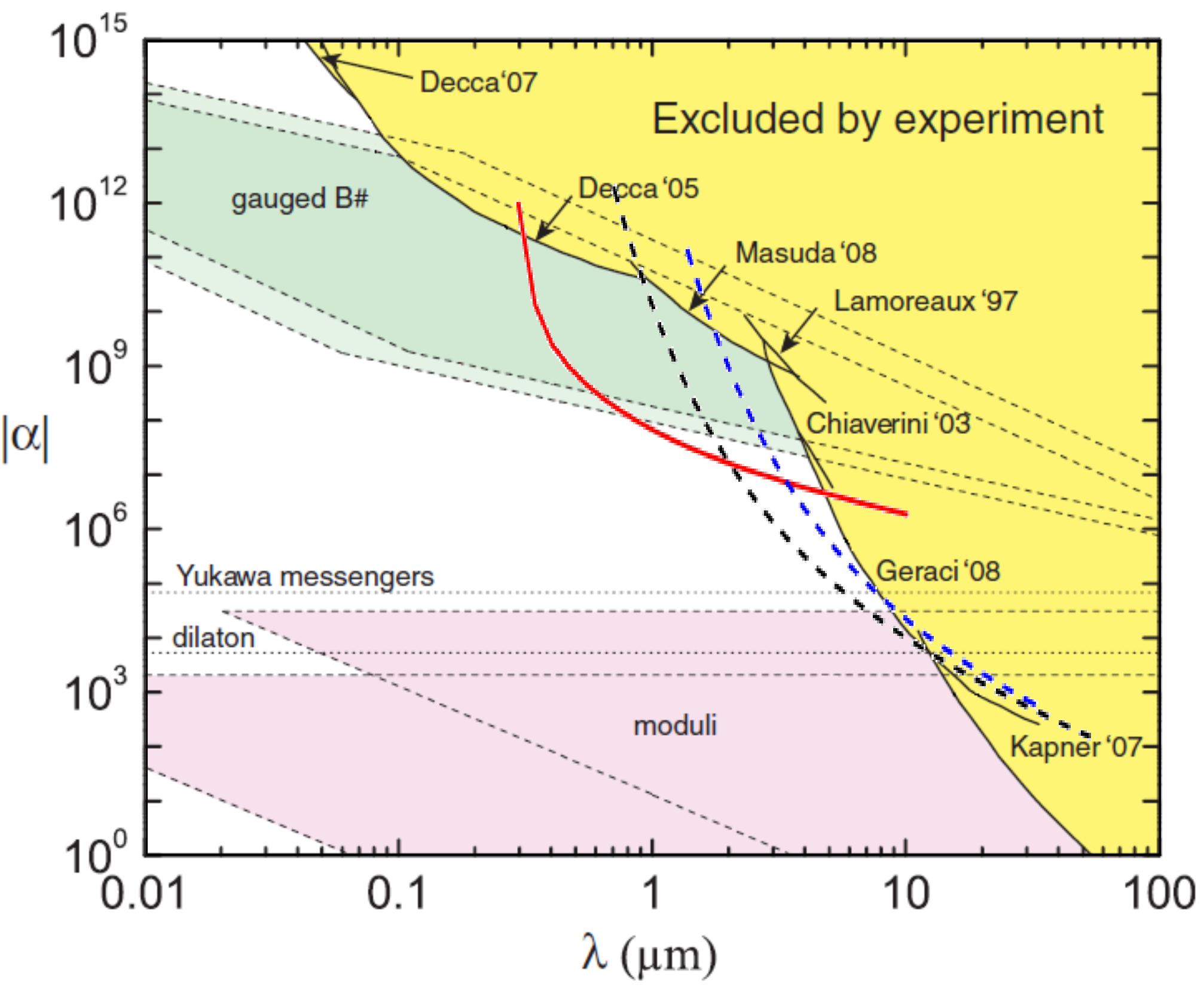}
\caption{(Color online) Present constraints (yellow region) and predicted constraints (coloured lines) for the FORCA-G experiment for the Yukawa potential in the $(\alpha_{\footnotesize{\text{Y}}},\lambda_{\footnotesize{\text{Y}}})$ plane. The colored zones are excluded. This figure is taken from \cite{MessinaPRA11} and adapted from \cite{GeraciPRL10}.}
\label{contraintes}
\end{figure}

After observing that the presence of the surface does not reduce drastically the lifetime of our modified Wannier-Stark states, we have to verify that this is not the case even in the presence of a Yukawa gravitational potential. Indeed, the main goal of the experiment FORCA-G is to search for a hypothetical deviation from Newton's law at short distance predicted by some unification theories. We have then to estimate the modification of the lifetimes of the states in the presence of such a deviation which can be written as an additional potential of the form \cite{MessinaPRA11}
\begin{equation}
U_{\footnotesize{\text{Y}}}=2\pi\alpha_{\footnotesize{\text{Y}}}G\rho_{\footnotesize{\text{s}}}m_a\lambda_{\footnotesize{\text{Y}}}^2e^{-2z/\lambda_{\footnotesize{\text{Y}}}},
\label{potyuk}
\end{equation}
where $G=6.67\times10^{-11}\,$m$^3\,$kg$^{-1}\,$s$^{-2}$ is the universal gravitational constant and $\rho_{\footnotesize{\text{s}}}$ is the density of our surface (for which we have chosen the density of the silicon $\rho_{\footnotesize{\text{s}}}=2.33\times10^{3}\,$kg\,m$^{-3}$). We want to stress here that the Yukawa potential presented in Eq. \eqref{potyuk} depends on two parameters $\alpha_{\footnotesize{\text{Y}}}$ and $\lambda_{\footnotesize{\text{Y}}}$. These parameters represent respectively the coupling strength of the non-Newtonian deviation and its typical range. The aim of experiments devoted to non-Newtonian gravitation is to impose constraints on the value of these two parameters. The present constraints and the ones predicted for the experiment FORCA-G are shown in Fig. \ref{contraintes}.

\begin{figure}\center
\includegraphics[height=5cm]{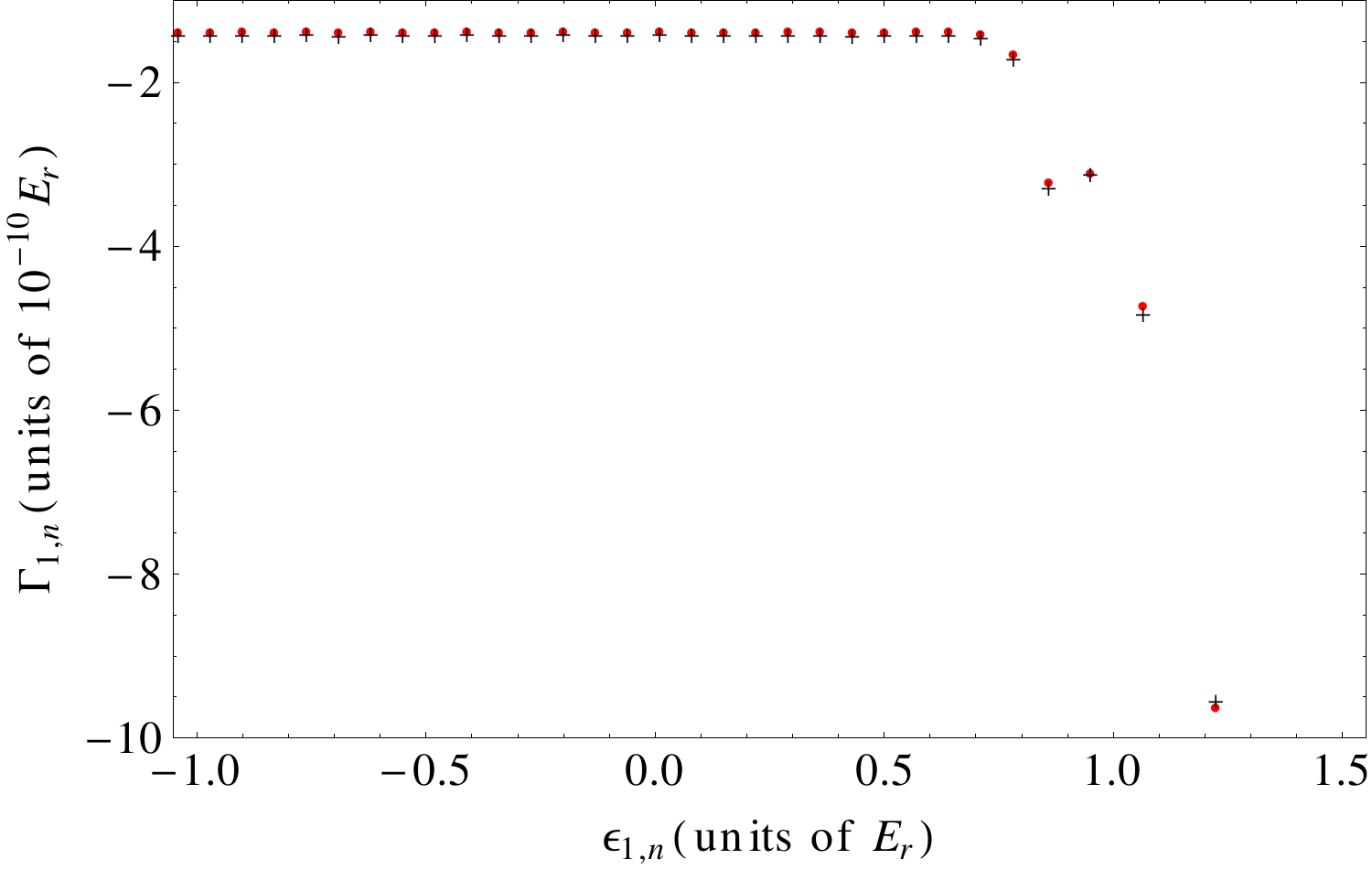}
\caption{(Color online) Values of the complex spectrum in the presence of a Yukawa potential (Red dots) compared with the modified Wannier-Stark complex spectrum (Black crosses). In this figure, we have chosen $U=2Er$, $\alpha_{\footnotesize{\text{Y}}}=10^{11}$ et $\lambda_{\footnotesize{\text{Y}}}=1\,\mu$m.}
\label{yukawa}
\end{figure}

In order to analyze the modification to the lifetimes induced by a hypothetical Yukawa term, we have to calculate the spectrum of the
Hamiltonian
\begin{equation}\begin{split}
H=&H_{\footnotesize{\text{WS}}}+U_{\footnotesize{\text{Y}}}\\
=&\frac{p^2}{2m_a}+\frac{U}{2}\left(1-\cos(2k_lz)\right)-m_agz\\
&+2\pi\alpha_{\footnotesize{\text{Y}}}G\rho_{\footnotesize{\text{s}}}m_a\lambda_{\footnotesize{\text{Y}}}^2e^{-2z/\lambda_{\footnotesize{\text{Y}}}}.
\end{split}\end{equation}
As done previously, we apply the transformation \eqref{scaling} to this Hamiltonian, obtaining the non-hermitian Hamiltonian
\begin{equation}\begin{split}
H^{(\theta)}=&\frac{e^{-2i\theta}p^2}{2m_a}+\frac{U}{2}\bigl(1-\cos(2k_lze^{i\theta})\bigr)\\
&-m_agze^{i\theta}+2\pi\alpha_{\footnotesize{\text{Y}}}G\rho_{\footnotesize{\text{s}}}m_a\lambda_{\footnotesize{\text{Y}}}^2e^{-2ze^{i\theta}/\lambda_{\footnotesize{\text{Y}}}}.
\end{split}\end{equation}
The complex spectrum resulting from this modified Hamiltonian is presented in Fig. \ref{yukawa}.
This figure shows that even if the presence of a deviation from Newton's gravitation law modifies the lifetimes of the atom in the trap, particularly for the closest wells, this lifetime remains very large in comparison with the duration of the experiment. Indeed, for practical reasons due to the coherence of the cold atoms in the experimental setup, the duration of the measurement in FORCA-G will not exceed $1\,$s. As a consequence, the Landau-Zener effect is not a limitation for the measurement of a deviation from Newton's law in our experiment. This can be more accurately verified from the analysis of Table \ref{lifeyuk}.
\begin{center}\begin{table}[h]\begin{tabular}{|c|c|c|c|}
\hline
\multicolumn{4}{|c|}{Far Regime}\\
\hline
$\lambda_{\footnotesize{\text{Y}}}$ ($\mu$m) & $\alpha_{\footnotesize{\text{Y}}}$ & $\Ima(E_{1,1}^{(\footnotesize{\text{Y}})})$ & $\tau^{(\footnotesize{\text{Y}})}$ ($10^{10}$\,s)\\
& & (units of $10^{-10}E_r$) & \\
\hline
0.70000 & 1.95007[12] & -6.25618 & 6.13778\\
0.80835 & 2.08378[11] & -8.09505 & 7.94185\\
0.93347 & 2.87073[10] & -8.44136 & 8.28161\\
1.07795 & 4.92177[9] & -8.51022  & 8.34916\\
1.24480 & 1.02282[9] & -8.52701  & 8.36564\\
1.43747 & 2.51801[8] & -8.53137  & 8.36991\\
1.65996 & 7.15759[7] & -8.53326  & 8.37177\\
1.91689 & 2.30712[7] & -8.53361  & 8.37211\\
2.21359 & 8.33762[6] & -8.53419  & 8.37268\\
2.55622 & 3.31836[6] & -8.53403  & 8.37252\\
\hline
\hline
\multicolumn{4}{|c|}{Near Regime}\\
\hline
$\lambda_{\footnotesize{\text{Y}}}$ ($\mu$m) & $\alpha_{\footnotesize{\text{Y}}}$ & $\Ima(E_{1,1}^{(y)})$ & $\tau^{(\footnotesize{\text{Y}})}$ ($10^{10}$\,s)\\
& & (units of $10^{-10}E_r$) & \\
\hline
0.10000 & 9.98047[11] & -8.55444 & 8.39255\\
0.11659 & 9.98047[11] & -8.55682 & 8.39488\\
0.13594 & 9.98047[11] & -8.55945 & 8.39746\\
0.15849 & 9.98047[11] & -8.56119 & 8.39917\\
0.18479 & 9.98047[11] & -8.56046 & 8.39845\\
0.21544 & 9.98047[11] & -8.55548 & 8.39357\\
0.25119 & 9.98047[11] & -8.54216 & 8.38050\\
0.29286 & 9.98047[11] & -8.51542 & 8.35427\\
0.34146 & 1.41907[10] & -8.53314 & 8.37165\\
0.39811 & 2.60450[9]  & -8.53385 & 8.37235\\
\hline\end{tabular}\caption{Table of the first ten imaginary parts of the complex energy in the first well as a function of the values of $\alpha_{\footnotesize{\text{Y}}}$ and $\lambda_{\footnotesize{\text{Y}}}$ calculated in \cite{MessinaPRA11}.}\label{lifeyuk}\end{table}\end{center}
This table confirms that the lifetimes of an atom in the trap are much longer than the duration of the measurement estimated in \cite{PelissonPRA12}.

\section{Discussion}\label{Sec:5}

As we have seen in the previous sections, the presence of a material surface as well as the assumptions of a deviation from the short-range gravitational law do not considerably limit the lifetimes of the metastable states of the trap. As a consequence, the assumption consisting in considering these states as pseudo-eigenstates remains valid under these assumptions. However, up to now, we have not taken into account the effect of the Casimir-Polder interaction between the quantized electromagnetic field in the presence of the surface and the atom. As we have seen in \cite{MessinaPRA11}, this effect is dominant at short distances inducing an important correction to the real energy levels of the atomic states. So we can suspect that it could have a non-negligible effect on the imaginary part of the spectrum as well.

As shown in \cite{MessinaPRA11}, the Casimir-Polder atom-surface interaction can be computed using second-order perturbation theory on the Hamiltonian interaction term. Using the same notation as \cite {MessinaPRA11}, the Hamiltonian describing our system can be expressed as
\begin{equation}\begin{split}
H&=H_0+H_\text{int}=H_\text{f}+H_\text{at}+H_{\footnotesize\text{WS}}+H_\text{int}\\
H_\text{f}&=\sumpk\,\hbar\omega\,\acpk\apk\\
H_\text{at}&=\hbar\omega_0\ket{e}\bra{e}\\
H_{\footnotesize\text{WS}}&=\frac{p^2}{2m_a}-m_agz+\frac{U}{2}\bigl(1-\cos(2k_lz)\bigr)\\
H_\text{int}&=-\bbm[\mu]\cdot\mathbf{\mathcal{E}}(\mathbf{r}).\end{split}\label{Htot}\end{equation}
The complete Hamiltonian is written as a sum of a term $H_0$ describing the free evolution of the atomic and field degrees of freedom. In particular, $H_\text{f}$ is the Hamiltonian of the quantum electromagnetic field, described by a set of modes $(p,\mathbf{k},k_z)$: here $p$ is the polarization index, taking the values $p=1,2$ corresponding to TE and TM polarization respectively, while $\mathbf{k}$ and $k_z$ are the transverse and longitudinal components of the wavevector. $H_\text{at}$ is the internal Hamiltonian of our two-level atom having ground state $\ket{g}$ and excited state $\ket{e}$ separated by a transition frequency $\omega_0$. While $H_\text{at}$ is associated to the internal atomic degrees of freedom, the term $H_{\footnotesize\text{WS}}$ accounts for the external atomic dynamics. The interaction between the atom and the quantum electromagnetic field is written here in the well-known multipolar coupling in dipole approximation \cite{PowerPhylTransRSocA59}, where $\bbm[\mu]=q\bbm[\rho]$ ($q$ being the electron's charge and $\bbm[\rho]$ the internal atomic coordinate) is the quantum operator associated to the atomic electric dipole moment and the electric field is calculated in the atomic position $\rv$.

In order to compute the correction to the complex energy spectrum due to the Casimir-Polder effect, we have used a non-hermitian equivalent of the complex perturbation theory based on a redefinition of the scalar product, which will be called \emph{c-product}, defined by \cite{Moiseyev2011}
\begin{equation}
\cbra{\varphi}\psi)=\int_{-\infty}^{+\infty}dx\,\varphi(x)\psi(x).
\label{cproduct}\end{equation}
Using this definition, we can define the correction term by the same way as in standard quantum mechanics and obtain the expression of the second order energy correction
\begin{equation}
\Delta E^{(2)}_n=\sum_{k\neq n}\frac{\bigl[\cbra{\psi^{(0)}_k}W^\theta\cket{\psi^{(0)}_n}\bigr]^2}{E_n^{(0)}-E_k^{(0)}},
\end{equation}
for a complex-scaled Hamiltonian of the form
\begin{equation}
H^\theta=H_0^\theta+W^\theta,
\end{equation}
where $H_0^\theta$ is the non-hermitian unperturbed Hamiltonian and $W^\theta$ is a small perturbation. We recall here that the first-order correction in the case of the dipolar approximation for the atom-field interaction is null due to the nature of the atomic dipolar operator $\bbm[\mu]$. After having scaled the Hamiltonian \eqref{Htot}, we are now prepared to calculate the complex correction knowing that the unperturbed states can be written as
\begin{equation}
\cket{\psi_{n,1}^{(0)}}=\ket{g}\ket{0_\epsilon(\mathbf{k},k_z)}\cket{n,1},
\end{equation}
where $n$ stands for the number of the well where the atom is initially placed (in the first Bloch band). We have to stress here that this state belongs to a product of three spaces and that the transformation \eqref{scaling} acts only on the external atomic states. As a consequence, we will use the \emph{c-product} only on the external atomic states whereas we keep the standard scalar product acting on the internal atomic states and field states. These assumptions lead to an expression for the second-order correction of the modified Wannier-Stark energy levels
\begin{widetext}
\begin{equation}
\Delta E^{(2)}_{m,1}=-\sumpk\sumnb\frac{\bigl[\cbra{\psi_{m,1}^{(0)}}H_{int}^\theta\ket{e}\ket{1_p(\mathbf{k},k_z)}\cket{n,b}\bigr]^2}{E_{n,b}^{(0)}-E_{m,1}^{(0)}+\hbar(\omega+\omega_0)},
\label{corr}\end{equation}\end{widetext}
where
\begin{equation}
H_\text{int}^\theta=-\bbm[\mu]\cdot\mathbf{E}(\mathbf{r}e^{i\theta}).
\end{equation}
Equation \eqref{corr} can be simplified considering that in a shallow trap, as in the case of FORCA-G, the Bloch bands above the second one are not trapped in the lattice so that the wavefunctions associated with these bands are very delocalized and the overlap with the states of the first band is very weak. This observation allows to restrict the sum over $b$ to its first two terms. Moreover, due to the extension of the modified Wannier-Stark wavefunctions in the first band, it is sufficient to take into account only 20 terms in the sum over $n$. Finally, the weakness of the imaginary part of the modified Wannier-Stark states as well as those of the difference of the real part of the Wannier-Stark spectrum with respect to the term $\hbar(\omega+\omega_0)$ allows us to ignore the contribution $E_{n,b}^{(0)}-E_{m,1}^{(0)}$ in the denominator in Eq. \eqref{corr}. All these simplifications lead to a new expression of the correction
\begin{widetext}\begin{equation}\begin{split}
\Delta E_{m,1}^{(2)}=&\int_0^{\infty}dz\,\int_0^{\infty}dz'\,\sum_{n=m-10}^{m+10}\Bigl[-\sumpk\Bigl(\Bigl[\frac{\cbra{m,1}z)(z\cket{n,1}\cbra{m,1}z')(z'\cket{n,1}}{\hbar(\omega+\omega_0)}\\
&+\frac{\cbra{m,1}z)(z\cket{n,2}\cbra{m,1}z')(z'\cket{n,2}}{\hbar(\omega+\omega_0)}\Bigr]\Alkzt\Aclkztp\Bigr)\Bigr]\\
=&-\int_0^{\infty}dz\,\int_0^{\infty}dz'\,\sumpk\sum_{n=m-10}^{m+10}\Bigl[\frac{WS_{m,1}(z)WS_{m,1}(z')}{\hbar(\omega+\omega_0)}\\
&\Alkzt\Aclkztp\left(WS_{n,1}(z)WS_{n,1}(z')+WS_{n,2}(z)WS_{n,2}(z')\right)\Bigr].
\end{split}\end{equation}\end{widetext}
where
$WS_{n,m}(x)$ state for the modified Wannier-Stark states taking into account the presence of the surface, and
\begin{equation}\begin{split}
\Alkzt&=-\bra{0_p(\mathbf{k},k_z)}\bbm[\mu]_{eg}\cdot\bbm[E](\mathbf{r}e^{i\theta})\ket{1_p(\mathbf{k},k_z)}\\
&=-\frac{i}{\pi}\sqrt{\frac{\hbar\omega}{4\pi\varepsilon_0}}e^{i\mathbf{k}\cdot\bbm[\rho]e^{i\theta}}\bbm[\mu]_{eg}\cdot\mathbf{f}_p(\mathbf{k},k_z,ze^{i\theta}),
\end{split}\end{equation}
where $\mathbf{f}_p(\mathbf{k},k_z,z)$ are the mode functions of the electric field in the presence of a perfectly conducting mirror in $z=0$ \cite{BartonJPhysB74}
\begin{equation}\begin{split}\mathbf{f}_1(\mathbf{k},k_z,z)&=\hat{\mathbf{k}}\times\hat{\mathbf{z}}\sin(k_zz)\\
\mathbf{f}_2(\mathbf{k},k_z,z)&=\hat{\mathbf{k}}\frac{ick_z}{\omega}\sin(k_zz)-\hat{\mathbf{z}}\frac{ck}{\omega}\cos(k_zz),\end{split}\end{equation}
being $\hat{\mathbf{k}}=\mathbf{k}/k$ and
$\hat{\mathbf{z}}=(0,0,1)$.
Unfortunately, the terms $\Alkzt$ and $\Aclkztp$ lead to a divergent integral when the coordinate $z$ is complex-scaled. Indeed, using the definitions of the electric field given in \cite{MessinaPRA11}, we obtain the following divergent expression for the correction
\begin{widetext}\begin{equation}\begin{split}
\Delta E_{m,1}^{(2)}=&-\frac{\hbar\bbm[\mu]_{eg}^2}{4\pi^3\epsilon_0}\int_0^{\infty}dz\,\int_0^{\infty}dz'\,\sumk\sum_{n=m-10}^{m+10}\\
&\Bigl\{\frac{W_{m,1}(z)W_{m,1}(z')}{\hbar(\omega+\omega_0)}\bigl(W_{n,1}(z)W_{n,1}(z')+W_{n,2}(z)W_{n,2}(z')\bigr)\\
&\omega\Bigl[\Bigl(1+\Bigl(\frac{ck_z}{\omega}\Bigr)^2\Bigr)\sin(k_zze^{i\theta})\sin(k_zz'e^{i\theta})+\left(\frac{ck}{\omega}\right)^2\cos(k_zze^{i\theta})\cos(k_zz'e^{i\theta})\Bigr]\Bigr\}.
\label{corrdiv}\end{split}\end{equation}\end{widetext} This
divergence is fundamental because the terms in $\cosh$ and $\sinh$ (coming from the $\sin$ and $\cos$ functions with complex arguments)
produce a divergence which is not sensitive to any ordinary regularization technique. This could be due to the fact that the atom is treated all along the calculation as a pointlike particle whereas its finite size should be taken into account. However, this effect is difficult to characterize because the scaling of the Hamiltonian makes the coordinate representing the atom-wall distance complex so that the atomic size should be defined as a complex quantity, which is far from natural. So, for the time being, the calculation of the effect of the Casimir-Polder interaction on the lifetimes of the atomic states remains an open problem.

\section{Conclusions}

In this paper we have studied the modifications to the lifetime of atoms trapped in an optical lattice in proximity of a surface. We have shown that the boundary condition introduced by the presence of the surface only marginally modifies the value of the ordinary Wannier-Stark lifetimes, leaving them almost infinite with respect to the duration of a measurement in a typical atomic-interferometry experiment such as the recently proposed FORCA-G. The same holds for the presence of a hypothetical Yukawa deviation from Newton's gravitational law. In our analysis, we have modeled the surface both as an infinite and a finite potential barrier, and considered both an atom above or below the surface.

As a natural development of our work, it could be interesting to investigate the behavior of the states in front of a more realistic surface, abandoning the assumptions of perfect conduction and infinite extension. This is the subject of ongoing work and will be part of an upcoming publication. As discussed in the last section, another problem which remains open is the precise calculation of the perturbation of the lifetimes due to the Casimir-Polder effect. However, we expect that this will not significantly reduce the lifetime. On the contrary, the Casimir-Polder force is attractive, so for atoms below the surface it will tend to counteract Landau-Zener tunnelling thus increasing the lifetime, rather than decreasing it. This assumption is made knowing that the Casimir-Polder potential is attractive toward the surface whereas the Landau-Zener effect tends to drag the atom away from the surface. Thus, at the moment we can assume that the Casimir-Polder effect should act on the {lifetimes in an opposite way with respect to} the Landau-Zener effect so that it should increase the lifetime of the atomic states. Finally, we point out that in this work we have not taken into account chemical processes between the atoms and the surface (atoms ``sticking'' to the surface) as we have considered the surface as a simple potential barrier. A more realistic analysis should include an additional potential describing that interaction, but is beyond the scope of the present paper.

\begin{acknowledgments}
This research is carried on within the project iSense, which acknowledges the financial support of the Future and Emerging Technologies (FET) programme within the Seventh Framework Programme for Research of the European Commission, under FET-Open grant number: 250072. We also gratefully acknowledge support by Ville de Paris (Emergence(s) program) and IFRAF. The authors thank N. Moiseyev, H. J. Korsch, Q. Beaufils, A. Hilico, B. Pelle and F. Pereira dos Santos for fruitful and stimulating discussions.
\end{acknowledgments}

\end{document}